\documentclass[preprint,aps]{revtex4}

\usepackage{graphicx}
\usepackage{dcolumn}
\usepackage{bm}
  
\begin{document}
\title{Comparing the Weighted Density Approximation with the LDA and GGA
for Ground State Properties of Ferroelectric Perovskites}
\author{Zhigang Wu}
\affiliation{Geophysical Laboratory, Carnegie Institution of Washington,
5251 Broad Branch Road, NW, Washington, DC 20015}
\author{R.E. Cohen}
\affiliation{Geophysical Laboratory, Carnegie Institution of Washington,
5251 Broad Branch Road, NW, Washington, DC 20015}
\author{D.J. Singh}
\affiliation{Center for Computational Materials Science,
Naval Research Laboratory, Washington, DC 20375}
\date{\today}

\begin{abstract}
First-principles calculations within the weighted density approximation (WDA) were
performed for ground state properties of ferroelectric perovskites
PbTiO$_3$, BaTiO$_3$, SrTiO$_3$, KNbO$_3$ and KTaO$_3$. We used the plane-wave 
pseudopotential method, a pair distribution function $G$ based on the uniform electron 
gas, and shell partitioning. Comparing with the local density approximation (LDA) and 
the general gradient approximation (GGA), we found that
the WDA significantly improves the equilibrium volume of these materials in 
cubic symmetry over both the LDA and GGA; 
Ferroelectric instabilities calculated by the WDA agree with the LDA and GGA very well;
At the experimental ferroelectric lattice, optimized atom positions by the WDA are 
in good agreement with measured data;
However the WDA overestimates the strain of tetragonal PbTiO$_3$ at experimental volume;
The WDA overestimates the volume of fully relaxed structures, but the GGA results are 
even worse. Some calculations were also done with other models for $G$. It is
found that a $G$ with longer range behavior yields improved relaxed structures. 
Possible avenues for improving the WDA are discussed.
\end{abstract}

\maketitle


\section{Introduction}
Since the early 1980s, first-principles calculations based on the density functional 
theory (DFT) have been implemented to compute diverse properties of 
piezoelectric and ferroelectric materials. The main difficulty within the DFT is how 
to treat the exchange-correlation (xc) energy accurately, because the exact form of it 
remains unknown. The local density approximation (LDA), in which the xc 
energy density ($\epsilon_{\rm xc}$) depends only on local charge density, dominated 
these calculations due to its simplicity and surprising success. When performed at 
the experimental volume, the LDA predicts many properties of ferroelectric 
materials, such as phonon frequencies, ferroelectric phase transitions, polarization, 
elasticity, etc., with extraordinary accuracy \cite{cohen99}. 
However it is well known that the LDA
overestimates the binding energy, and underestimates the bond length by 1-2\%, which 
results in the calculated equilibrium volume normally 3-6\% less than experiment. On 
the other hand, ferroelectric properties are extremely sensitive to volume. For 
example, the ferroelectric instabilities in BaTiO$_3$ \cite{cohen90} and KNbO$_3$ 
\cite{singh92,postn93,yu95,singh95} are severely reduced, if not totally eliminated, 
at the LDA zero pressure volume. Even at the experimental volume, the LDA  
incorrectly predict certain properties, e.g., it overestimates the strain of the tetragonal
PbTiO$_3$ \cite{cohen99} and the static dielectric constant $\varepsilon _{\infty }$ 
\cite{corso94,resta97}, and it underestimates the band gap \cite{hamann79,yin82}.

The generalized gradient approximation (GGA) \cite{lang83,perdew96}, which
includes the density gradients, is the natural next step beyond the LDA. Generally 
speaking, the semi-local GGA tends to improve upon the LDA 
in many aspects, especially for atomic energies and structural energy differences 
\cite{perdew96,perdew92a,hammer93}. For ferroelectric properties and band gaps, the GGA 
normally predicts very similar results to the LDA at experimental volume.
However the GGA tends to overestimate the bond length by about 1\% \cite{filip94, singh95}.
To illustrate this we present a full relaxation of tetragonal $P4mm$ PbTiO$_3$ 
within both the LDA (Hedin-Lundqvist \cite{hedin71}) and the GGA 
(The PBE \cite{perdew96} version 
was used throughout this paper.) using the linearized augmented planewave 
method with local orbital extensions (LAPW+LO) \cite{singh94}. 
As seen in Table \ref{tab1}, at the experimental volume, the GGA predicts a strain 
better than the LDA, but the fully relaxed GGA structure is much worse than that of 
the LDA. We also used the ABINIT package \cite{abinit1,abinit2}, which is 
based on pseudopotentials and a planewave basis. For the fully relaxed tetragonal 
PbTiO$_3$, the LDA predicts a volume of 60.34 \AA$^3$, and $c/a = 1.042$, while the 
GGA gives a volume of 70.54 \AA$^3$, and $c/a = 1.24$. The ABINIT results 
are in excellent agreement with the LAPW results.
The failures of the LDA and GGA indicate that more 
complicated non-local approximations are needed.

The weighted density approximation (WDA) \cite{gunna76,alonso78,gunna79}, within which
$\epsilon_{\rm xc}$ depends on charge density over a finite region, was advanced
in the late 1970's. The WDA assumes that any inhomogeneous electron gas can be regarded 
as continuously being deformed from a homogeneous electron gas, so the real 
pair-distribution function of the inhomogeneous gas can be replaced by that of 
a homogeneous gas at every point with certain weighted density. By constructing 
a model xc hole, the sum rule is fulfilled. One problem that both the LDA and the 
GGA have is the self-interaction error, i.e., the imperfect cancellation of Hartree 
and xc terms in the one-electron limit. The self-interaction-correction (SIC) method 
\cite{zunger81} predicts band gaps of transition-metal oxides in good agreement with 
experiments \cite{svane90}. It demonstrates the importance of SIC.
However the SIC method is not efficient because its potential is orbital-dependent.
The WDA is free of self-interaction for the one-electron limit, and its potential
is directly a functional of the charge density, and not orbital dependent.
For the non-local (uniform) limit, it  
reduces to the LDA. It has been reported that the WDA significantly
improves atomic energies \cite{gunna79,grit93} and the equilibrium volume of some simple
bulk solids \cite{singh93} over the LDA. It is promising to apply the WDA for
ground state properties of more complicated ferroelectric perovskites. 

The selected perovskites PbTiO$_3$, BaTiO$_3$, SrTiO$_3$, KNbO$_3$ and KTaO$_3$
are both technically and theoretically important ferroelectric materials. They are 
also prototypes of more complicated and interesting ferroelectric relaxors such as 
($1-x$)Pb(Mg$_{1/3}$Nb$_{2/3}$)O$_{3}$-$x$PbTiO$_3$ (PMN-PT) and 
($1-x$)Pb(Zn$_{1/3}$Nb$_{2/3}$)O$_{3}$-$x$PbTiO$_3$ (PZN-PT). They have previously been
studied extensively to understand ferroelectricity 
\cite{cohen90,singh92,postn93,yu95,singh95,singh96,lasota}.
In this paper, first a brief overview of the WDA formalism will be presented in 
section II. Then we will report the WDA results of ground state properties of these 
perovskites in section III. In section IV, we discuss avenues for improving the
WDA focussing on the symmetry under particle exchange in the
pair correlation function.

\section{Formalism}

The general form of the xc energy in the DFT scheme can be expressed as
(All equations are in atomic units, $\hbar = m = e^2 = 1$.)
\begin{equation}
E_{\rm xc}[n]= \frac{1}{2} \int n({\bf r}) d{\bf r}
\int \frac{\bar{n}_{\rm xc}({\bf r}, {\bf r}^{\prime})}
{|{\bf r-r}^{\prime}|}d{\bf r}^{\prime},
\label{eq2.1}
\end{equation}
here the xc hole density $\bar{n}_{\rm xc}({\bf r}, {\bf r}^{\prime})$ is defined as 
\begin{eqnarray}
\bar{n}_{\rm xc}({\bf r}, {\bf r}^{\prime}) =
   n({\bf r}^{\prime}) \int^{1}_{0} 
    [\bar{g}^{n}_{\rm xc}({\bf r}, {\bf r}^{\prime}; \lambda) - 1] d \lambda   
   \nonumber \\
   \equiv n({\bf r}^{\prime})
          [\bar{g}^{n}_{\rm xc}({\bf r}, {\bf r}^{\prime}) - 1],
\label{eq2.2}
\end{eqnarray}
where $\bar{g}^{n}_{\rm xc}({\bf r}, {\bf r}^{\prime})$
is the coupling-constant averaged pair-distribution function of a system with a 
density $n({\bf r})$ \cite{gunna1976}. This function is symmetric
\begin{equation}
\bar{g}^{n}_{\rm xc}({\bf r}, {\bf r}^{\prime}) =
\bar{g}^{n}_{\rm xc}({\bf r}^{\prime}, {\bf r}),
\label{eq2.3}
\end{equation}
and the xc hole satisfies the following sum rule
\begin{equation}
\int \bar{n}_{\rm xc}({\bf r}, {\bf r}^{\prime})d{\bf r}^{\prime} = -1.
\label{eq2.4}
\end{equation}
Although the function $\bar{g}_{\rm xc}$ of a uniform gas is known
with high accuracy based on Monte Carlo simulations \cite{perdew92b}, the 
exact form of $\bar{g}^{n}_{\rm xc}({\bf r}, {\bf r}^{\prime})$ is still elusive
for an arbitrary inhomogeneous system. Different density approximations
make different approximations to $\bar{g}_{\rm xc}$. In the WDA approach  
$\bar{g}^{n}_{\rm xc}({\bf r}, {\bf r}^{\prime})$ is approximated by a 
model function $G$ (e.g., a uniform type),
\begin{equation}
G[|{\bf r-r}^{\prime}|, \bar{n}({\bf r})] = 
\bar{g}^{n}_{\rm xc}({\bf r}, {\bf r}^{\prime}) - 1,
\label{eq2.5}
\end{equation}
where the parameter $\bar{n}({\bf r})$ is the weighted density, and it can be 
determined from the sum rule
\begin{equation}
\int n({\bf r}^{\prime}) G[|{\bf r-r}^{\prime}|, \bar{n}({\bf r})]
d{\bf r}^{\prime} = -1.
\label{eq2.6}
\end{equation}
By definition, the function $G$ is not symmetric
\begin{equation}
G({\bf r}, {\bf r}^{\prime}) \neq
G({\bf r}^{\prime}, {\bf r}).
\label{eq2.7}
\end{equation}
Since $\bar{n}({\bf r})$ is only a function of ${\bf r}$, not ${\bf r}^{\prime}$,
it leads to wrong asymptotic behavior at the low density limit \cite{gunna79}, which we will
discuss in section IV. 

The natural choice of $G$ would be obtained from $g$ for a uniform electron gas, but
since the systems of interest are inhomogeneous, other types of model function $G$
could be better\cite{gunna80,mazin98}. Among them, the Gunnarsson-Jones (GJ \cite{gunna80}) 
ansatz ensures the correct behavior at large distance
\begin{equation}
G^{\rm GJ}(r,n) = c_1(n)\{1-{\rm exp}(-[\frac{r}{c_2(n)}]^{-5})\},
\label{eq2.8}
\end{equation}
while Rushton {\it et al.\ } (RTZ \cite{rush2002}) used a simple Gaussian function 
\begin{equation}
G^{\rm RTZ}(r,n) = c_1(n){\rm exp}(-[\frac{r}{c_2(n)}]^{2}),
\label{eq2.9}
\end{equation}
which resembles the exact one for the short distance limit, and it is also a good 
approximation to the uniform $G$. The parameters $c_1$ and $c_2$ can be determined 
from the following conditions
\begin{equation}
n \int G(r,n) d^3 r = -1,
\label{eq2.10} 
\end{equation}
\begin{equation}
n \int \frac{G(r,n)}{r} d^3 r = \epsilon^{\rm hom}_{\rm xc} (n),
\label{eq2.11}
\end{equation}
where $\epsilon^{\rm hom}_{\rm xc} (n)$ is the xc energy density of a homogeneous gas
with density $n$.

The WDA xc energy is
\begin{equation}
E^{\rm WDA}_{\rm xc}[n] = 
  \frac{1}{2} \int\int \frac{n({\bf r})n({\bf r}^{\prime})}{|{\bf r-r}^{\prime}|}
  G[|{\bf r-r}^{\prime}|, \bar{n}({\bf r})] d{\bf r} d{\bf r}^{\prime},
\label{eq2.12}
\end{equation}
and the corresponding xc potential $v_{\rm xc}({\bf r})$ is the functional 
derivative of $E_{\rm xc}$,
\begin{equation}
v^{\rm WDA}_{\rm xc}({\bf r}) = 
     \frac{\delta E^{\rm WDA}_{\rm xc}[n]}{\delta n({\bf r})} = 
     v_{1}({\bf r}) + v_{2}({\bf r}) + v_{3}({\bf r}),
\label{eq2.13}
\end{equation}
where
\begin{equation}
v_{1}({\bf r}) =  \frac{1}{2} \int \frac{n({\bf r}^{\prime})}{|{\bf r-r}^{\prime}|}
          G[|{\bf r-r}^{\prime}|, \bar{n}({\bf r})] d{\bf r}^{\prime}
          = \epsilon^{\rm WDA}_{\rm xc}({\bf r}),
\label{eq2.14}
\end{equation}
\begin{equation}
v_{2}({\bf r}) = \frac{1}{2}  \int \frac{n({\bf r}^{\prime})}{|{\bf r-r}^{\prime}|}
     G[|{\bf r-r}^{\prime}|, \bar{n}({\bf r}^{\prime})] d{\bf r}^{\prime},
\label{eq2.15}
\end{equation}
\begin{equation}
v_{3}({\bf r}) =  
     \frac{1}{2} \int\int \frac{n({\bf r}^{\prime})n({\bf r}^{\prime\prime})}
     {|{\bf r}^{\prime} -{\bf r}^{\prime\prime}|}
     \frac{\delta G[|{\bf r}^{\prime} -{\bf r}^{\prime\prime}|, 
     \bar{n}({\bf r}^{\prime})]}{\delta n({\bf r})} 
     d{\bf r}^{\prime} d{\bf r}^{\prime\prime}.
\label{eq2.16}
\end{equation}
Examination of the above suggests that implementation of $E^{\rm WDA}_{\rm xc}$ 
and $v^{\rm WDA}_{\rm xc}({\bf r})$ could be cumbersome. However, in a plane-wave 
representation, by using the convolution theorem, these terms can be evaluated 
efficiently, as detailed in Ref. \cite{singh93}.

One subtle issue in the WDA implementation is shell partitioning \cite{gunna79}. The
WDA scheme does not describe the exchange interaction between core and valence states
very well because of the use of local $\bar{n}({\bf r})$ in the model function $G$. 
Physically the range of integration of $G$ is similar to the size of atoms, and there 
is no distinction between core and valence electrons to the contribution of 
$E^{\rm WDA}_{\rm xc}$. As a result core and valence electrons dynamically 
screen valence electrons equally, which leads to non-zero exchange energy between
core and valence states outside core regions. On the other hand, the 
LDA can give correct inter-shell contributions, since the LDA depends only on local 
density, and the core density vanishes outside core regions. Based on this 
observation, a shell partitioning approach was proposed \cite{gunna79,singh93}, in 
which the valence-valence interactions are treated with the WDA, while core-core 
and core-valence interactions with the LDA. The total xc energy is written 
\begin{equation}
E_{\rm xc}[n] = E^{\rm LDA}_{\rm xc}[n] 
+ E^{\rm WDA}_{\rm xc}[n_v] - E^{\rm LDA}_{\rm xc}[n_v],
\label{eq2.17}
\end{equation}
where $n_{\rm v}$ is the valence density and $n$ is the total density.
The sum rule becomes
\begin{equation}
\int \{n_{\rm v}({\bf r}^{\prime}) G[|{\bf r-r}^{\prime}|,\bar{n}({\bf r})]
+ n_{\rm c}({\bf r}) G[|{\bf r-r}^{\prime}|,n({\bf r})]\} d{\bf r}^{\prime} = -1,
\label{eq2.18}
\end{equation}
where $n_{\rm c}$ is the core density. Ref \cite{wu2003} shows why this sum rule
must be used instead of a simpler one 
$\int \{n_{\rm v}({\bf r}^{\prime}) G[|{\bf r-r}^{\prime}|,\bar{n}({\bf r})] = -1$.
The corresponding $v_{\rm xc}$ for core and valence states  will be different 
since $E_{\rm xc}$ depends explicitly on both core and valence density. The detailed 
derivation and formulas can be found in the appendix of Ref. \cite{wu2004}. We can still
use the Hellmann-Feynmann theorem \cite{hf37} to determine atomic forces, and we 
found that the directly calculated forces agree with the numerical results of finite 
energy difference very well with the WDA shell partitioning.

\section{Results}
\subsection{Technical details}

The WDA was implemented \cite{singh93} within a plane-wave basis pseudopotential 
method. The pseudopotentials are of the hard Troullier-Martins type \cite{tm91}. The 
semi-core states of metal ions include $3s$ and $3p$ states of K and Ti, $4s$ and $4p$ 
states of Nb and Sr, $5s$ and $5p$ states of Ta and Ba, and $5d$ states of Pb. $2s$ 
states of O are also treated as semi-core. We used the WDA for valence states, the 
LDA for semi-core states, and pseudized lower states. Plane-wave basis sets with cut-off 
of 132 Ry were tested and found to be highly converged. A $4 \times 4 \times 4$ ${\bf k}$-point 
mesh was exploited except for frozen-phonon soft-mode calculations of rhombohedral 
BaTiO$_3$ and KNbO$_3$, where a denser $6 \times 6 \times 6$ ${\bf k}$-mesh was used 
because the ferroelectric double well depth
is only about 2-3 mRy. The standard 3-point interpolation was employed to obtain 
$\bar{n}({\bf r})$ with a logarithmic grid of increment $\bar{n}_{i+1}=1.25\bar{n}_{i}$.
We tested the convergence by the 6-point interpolation which makes negligible difference. 
For the function $G$, we used a uniform type \cite{ggp02} for the following calculations. In 
the last part of this section we will also present results with other types of $G$.

\subsection{Equilibrium volume of cubic structures}

We first calculated the lattice constant of these materials constrained in
cubic symmetry. The experimental values in Table \ref{tab2} are those
extrapolated to $T=0$. The zero point corrections are expected to be less than
0.3\%. As mentioned before, the LDA lattice constant is about 1-2\% less 
than experiment, and this small error is big enough to make many ferroelectric 
properties incorrect. On the other hand, the GGA results are better than the LDA, but
they are a little too large for BaTiO$_3$, SrTiO$_3$, and KTaO$_3$.
The WDA dramatically improves the lattice constant over the LDA, and it is also
better than the GGA. Actually all these WDA lattice constants are very close to 
experimental data except for PbTiO$_3$.  At low temperature PbTiO$_3$ is 
tetragonal with a big strain of 6\%, and other four perovskites are very close 
to the cubic structure. The extrapolation of the high temperature cubic
data of PbTiO$_3$ therefore is not expected to be as reasonable as the other cases.

\subsection{Ferroelectric instabilities}

\begin{figure}
\includegraphics[width=0.45\linewidth]{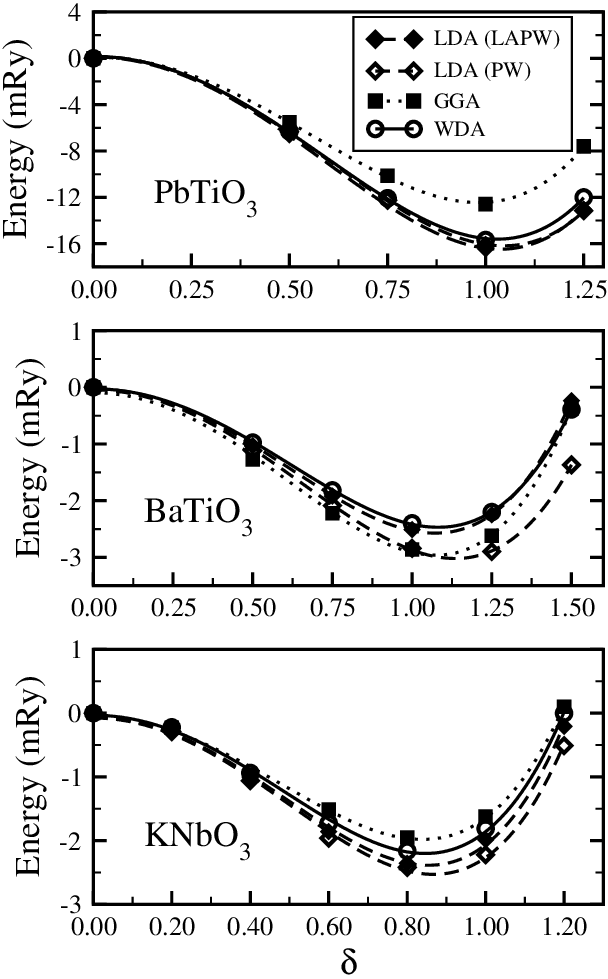}
\caption{\label{fig1}
Total energy as a function of soft-mode displacement in tetragonal 
PbTiO$_3$ ($c/a=1.063$), rhombohedral BaTiO$_3$ and KNbO$_3$ with the 
LDA, GGA, and WDA (uniform electron gas $G$). Here $\delta$ is 
the displacement relative to experiment.}
\end{figure}

We studied the ferroelectric instability in tetragonal PbTiO$_3$ and 
rhombohedral BaTiO$_3$ and KNbO$_3$. We displaced atom positions according to 
the experimental soft-mode distortion patterns at the experimental lattice.
For tetragonal PbTiO$_3$ $c/a=1.063$, while for rhombohedral BaTiO$_3$ and 
KNbO$_3$, we neglected the tiny lattice distortion from the cubic structure. 
As mentioned before, the LDA and GGA describes the ferroelectric 
instability very well at the experimental structure, so one may hope that the WDA 
retains this good feature. Fig. \ref{fig1} shows the calculated curves of
energy versus ferroelectric displacement along [001] for PbTiO$_3$ and [111] for
BaTiO$_3$ and KNbO$_3$. In all cases, the WDA curves match with the LDA ones 
very well. It shows that the WDA can predict ferroelctric phase transitions
as well as the LDA. On the other hand, the GGA predicts a smaller energy difference 
between paraelectric and ferroelectric states of PbTiO$_3$ than 
the LDA, but agrees well with the LDA for BaTiO$_3$ and KNbO$_3$.

\subsection{Structural optimization}
Since there is no difficulty calculating atomic forces with the WDA approach, we performed
structural optimization for tetragonal $P4mm$ PbTiO$_3$ and rhombohedral $P3m1$  
BaTiO$_3$ and KNbO$_3$. First we only optimized internal parameters at experimental
lattices, and the relaxed atomic positions are shown in Tables \ref{tab3}, \ref{tab4}
and \ref{tab5} for PbTiO$_3$, BaTiO$_3$ and KNbO$_3$ respectively. In all cases, the 
LDA, GGA, and WDA predict very similar results which are in good agreement 
with experiment except for KNbO$_3$ whose theoretical displacements from ideal 
positions are about 30\% less than measured data \cite{hewat}. The consistency
of theoretical results suggests re-examination of these experimental data.

We also did full relaxations for the above materials. At the experimental volume, 
the WDA predicts $c/a = 1.106$ for PbTiO$_3$, which is close to the LDA 
value of 1.112, while the GGA gives $c/a = 1.068$. Compared with the measured 
value of 1.063, the WDA is not as good as the GGA in this case. 
For the fully relaxed structure, the WDA predicts a large $c/a=1.19$ and a volume 8.2\%
bigger than experiment, which are poorer than the LDA, but still a little better 
than the GGA. To demonstrate that this is not because of poor pseudopotentials or
other problems in the planewave method, we compared the energy difference between 
the LDA equilibrium structure and the WDA equilibrium structure. For the LDA, the 
planewave code predicts -11.8 mRy, while the LAPW gives -12.3 mRy. It proves 
that the planewave method is reliable. On the other hand, the WDA energy 
difference is 9.4 mRy, which is similar in magnitude with opposite sign.
We also calculated the equilibrium structures of rhombohedral BaTiO$_3$ and KNbO$_3$, as 
shown in Table \ref{tab6}. We can conclude that the LDA underestimates the fully
relaxed volume by almost the same amount as constrained in the cubic structure, 
while the WDA overestimates it, and the GGA overestimates it even more.

\subsection{Results with other $G$ functions}
\begin{figure}
\includegraphics[width=0.45\linewidth]{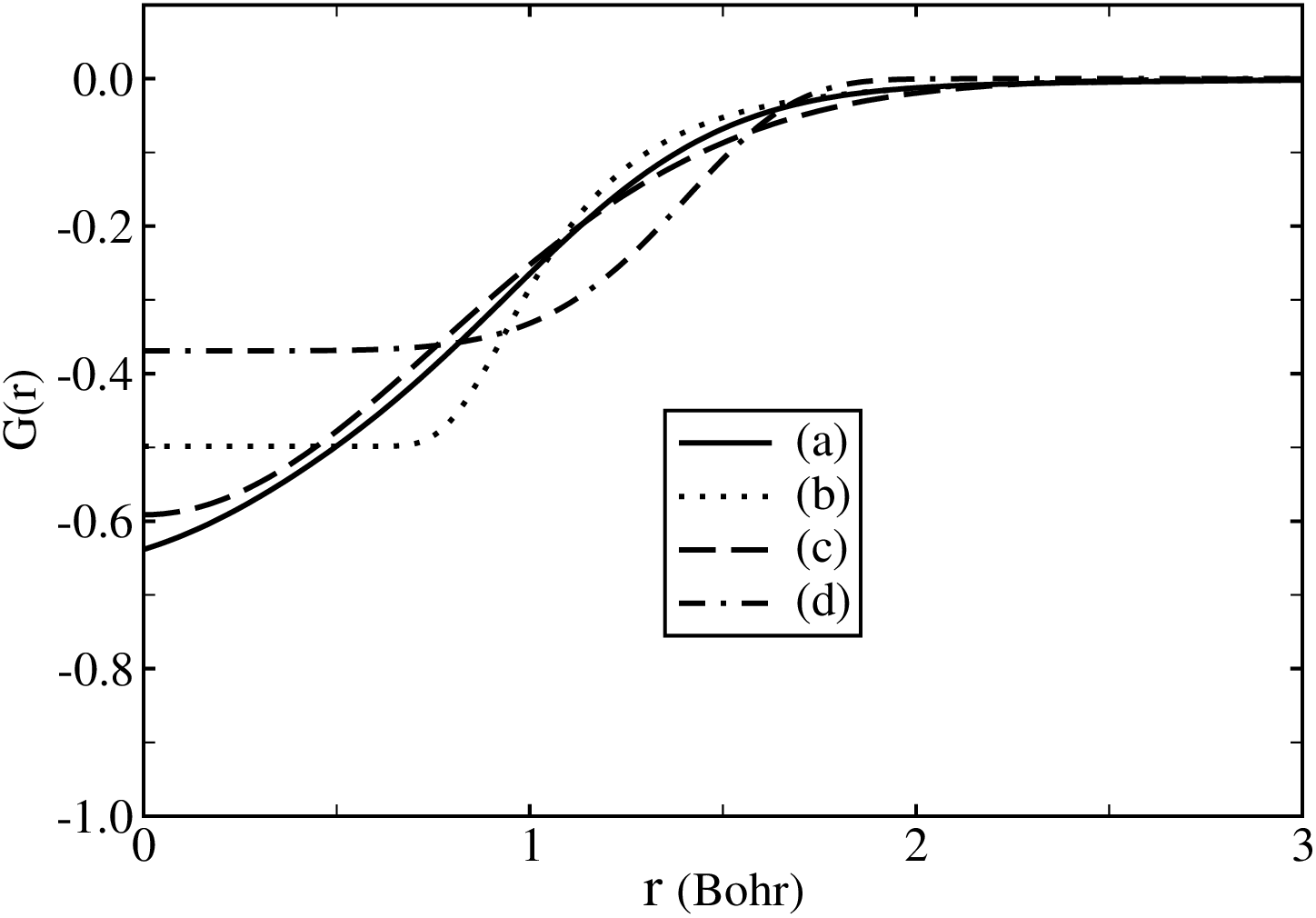}
\caption{\label{fig2} Four types of function $G(r,n)$ 
with $r_{s}=1.0$ ($n = 0.2387$). (a), (b), (c) and (d) 
correspond to the types defined in the last subsection of 
section III. }
\end{figure}

Because there is no reason for the uniform $G$ to be the best choice, we also tried other 
types of function $G$. We denote the uniform type as (a), the GJ type as (b), the simple 
Gaussian (RTZ) type as (c), and type (d) as $G(r,n) = c_1(n){\rm exp}
(-[\frac{r}{c_2(n)}]^{6})$. To clearly visualize them, we show these functions together
at $n=1.0$ in Fig. \ref{fig2}. The simple Gaussian type is a good 
approximation to the uniform $G$, the type (d) has longer interaction range than others,
and the GJ type agrees with the uniform one when $r > 1.745$. 

We recalculated the equilibrium lattice constants of the cubic structure with the 
other three $G$s. As shown in Table \ref{tab7}, the newly calculated lattice constants 
are very similar. Compared with the previous (uniform $G$) results, the GJ type 
predicts slightly larger lattice constants, while the simple Gaussian form and type 
(d) predict slightly smaller ones.

However the choice of $G$ is more sensitive for the fully relaxed structure. Comparing with
the uniform $G$, We found that the GJ type predicts even worse results, the simple 
Gaussian gives very similar results, and the type (d) is particularly better. 
For tetragonal PbTiO$_3$ at experimental volume, its optimized $c/a$ becomes 1.078, and
for the fully relaxed structure $c/a$ is 1.092 and the volume is only 0.8\% larger
than experiment. However, as listed in Table \ref{tab8}, the volumes of rhombohedral 
BaTiO$_3$ and KNbO$_3$ do not improve as much as that of tetragonal PbTiO$_3$, they are
still 2.4\% and 1.3\% larger than experiment respectively.

\begin{figure}
\includegraphics[width=0.6\textwidth]{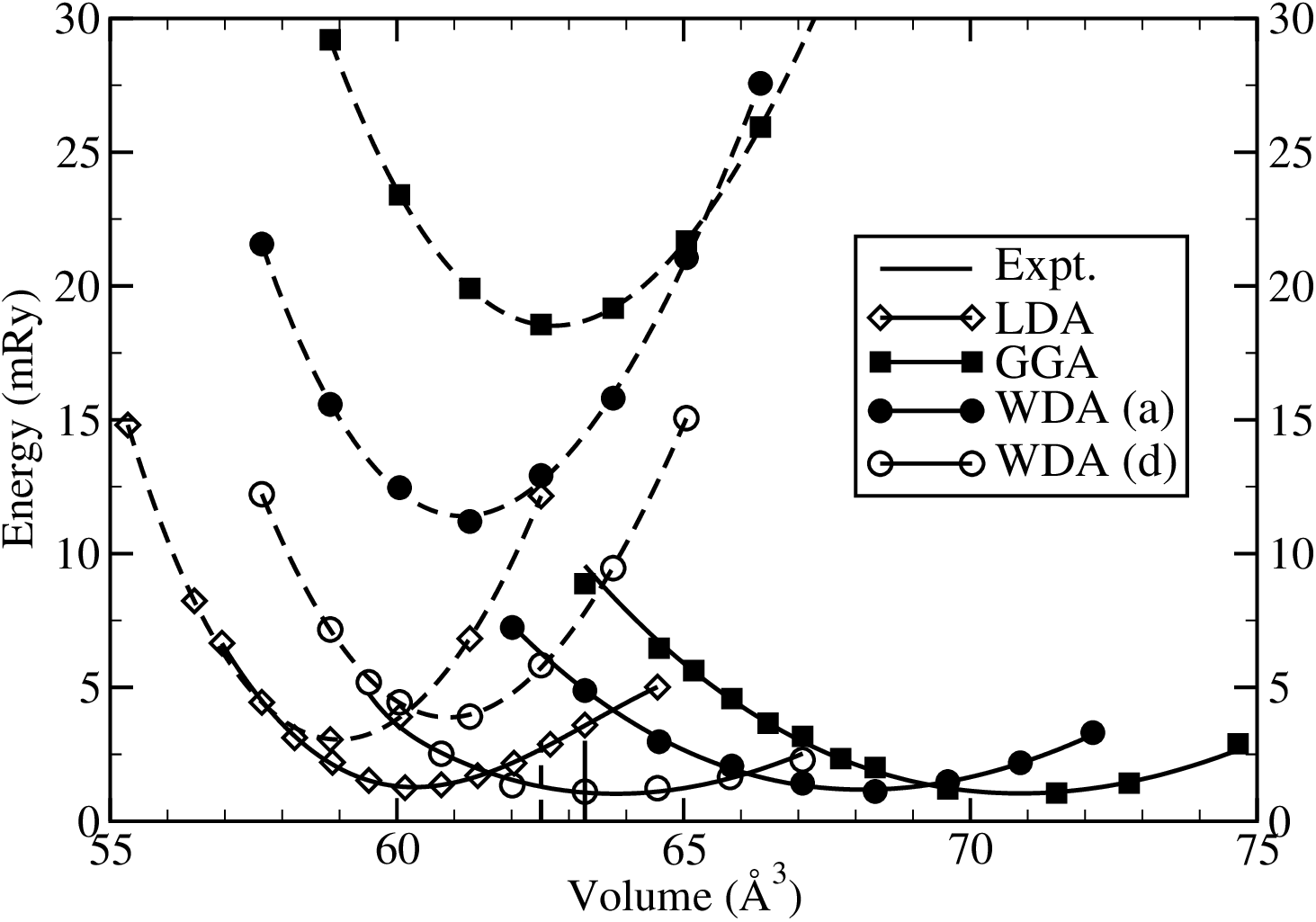}
\caption{\label{fig3} Energy as a function of volume for the 
cubic and the relaxed tetragonal PbTiO$_3$, with the LDA, GGA, 
and WDA (two types of $G$). Dashed lines are cubic structures, 
and solid lines are tetragonal structures.} 
\end{figure}

We could further improve the WDA results by tuning the shape of function $G$, but it is
hard to justify, and also it is difficult to make a single $G$ to fit all properties of 
all materials. We need a simple and reasonable $G$, such as the uniform one, which can
predict good ground state properties for both the cubic and the fully relaxed structures.
Actually since different $G$s predict very similar equilibrium volumes
for the cubic structure, we expect their results for the fully relaxed
structure should also be similar to the cubic structure, just like the LDA. 
So we focus on the WDA method itself. If one draws energy 
versus volume curves for the cubic and optimized tetragonal PbTiO$_3$ as shown in 
Fig. \ref{fig3}, one can see that the WDA curves for the cubic structure have 
similar curvatures to the LDA one, while for the
tetragonal structure the WDA curves are too flat on the right side in comparison with 
the LDA. It means that for the relaxed structure the WDA predicts an energy increase 
smaller than expected on the large volume side. This suggests an asymptotic 
problem, as is discussed in the next section.

\section{Discussions and Prospects}
\subsection{Asymptotic behavior}
An accurate functional approximation should fulfill at least some of the following
conditions: (1) sum rule of the xc hole; (2) for a slowly varying density, it should
recover the uniform gas limit; (3) absence of self-interaction; 
(4) asymptotic properties of xc energy and potential.
One may verify that from the exact DFT expression (Eq. \ref{eq2.1}) the xc energy 
density $\epsilon_{\rm xc}({\bf r})$ far away from the nucleus (low density limit) are
\begin{equation}
\lim_{r\to\infty} \epsilon_{\rm xc}({\bf r}) \to - \frac{1}{2r}.
\label{eq4.1}
\end{equation}
Since $g^{n}_{\rm xc} ({\bf r},{\bf r}^{\prime})$ is symmetric, one can derive that
\begin{equation}
v_{\rm xc}({\bf r}) = 2 \epsilon_{\rm xc}({\bf r}) + v_3({\bf r}),
\label{eq4.2}
\end{equation}
and because 
$\frac{\delta g^{n}_{\rm xc}({\bf r}^{\prime}, {\bf r}^{\prime \prime})} {\delta n({\bf r})}$
vanishes exponentially in the above limit,
\begin{equation}
\lim_{r\to\infty} v_{\rm xc}({\bf r}) \to - \frac{1}{r}.
\label{eq4.3}
\end{equation}

The asymptotic conditions \ref{eq4.1} and \ref{eq4.3} are the consequence of requirement
of cancellation of self-interaction in the Hartree terms. They look simple, but are very 
difficult to fulfill simultaneously. One can easily construct an energy or a potential 
satisfying the above conditions separately, but difficulty arises in making 
$v_{\rm xc}({\bf r})$ as the functional derivative of $E_{\rm xc}[n({\bf r})]$. 
In the LDA and GGA, both $\epsilon_{\rm xc}({\bf r})$ and 
$v_{\rm xc}({\bf r})$ go to zero exponentially for large $r$. 
Asymptotically they are less attractive than they should be. In the WDA, 
$\epsilon^{\rm WDA}_{\rm xc}({\bf r})$ satisfies condition \ref{eq4.1},
but the unsymmetrical function $G$ (Eq. \ref{eq2.7}) leads to 
$v_{1}({\bf r}) \neq v_{2}({\bf r})$, and 
$v_{2}({\bf r})$ goes to zero exponentially also, giving that 
$v^{\rm WDA}_{xc}({\bf r})$ at large $r$ to be the same as 
$\epsilon^{\rm WDA}_{\rm xc}({\bf r})$
\begin{equation}
\lim_{r\to\infty} v^{\rm WDA}_{\rm xc}({\bf r}) \to - \frac{1}{2r},
\label{eq4.4}
\end{equation}
which is off by a factor of 1/2.

The correct asymptotic behavior of $\epsilon^{\rm WDA}_{\rm xc}({\bf r})$
is due to the correct handling self-interaction in the WDA, while the wrong 
asymptotic behavior of $v^{\rm WDA}_{\rm xc}({\bf r})$ results from the 
violated symmetry of G under exchange $({\bf r},{\bf r}^{\prime}) 
 \longleftrightarrow ({\bf r}^{\prime},{\bf r})$. 
In order to overcome this problem so that the WDA 
will behave closer to the exact DFT and eventually it will circumvent
its failure for the fully relaxed structure, we can symmetrize $G$.

\subsection{A new approach: symmetrization of the WDA}
A symmetric $G$ satisfying Eq. \ref{eq2.3} will make both 
$\epsilon^{\rm WDA}_{\rm xc}({\bf r})$ and $v^{\rm WDA}_{\rm xc}({\bf r})$
have the correct asymptotic behavior. The simplest approach is to make $G$ 
depends on the weighted density at both ${\bf r}$ and ${\bf r}^{\prime}$, for 
example, a sum form would be
\begin{equation}
G^{\rm SWDA}({\bf r}, {\bf r}^{\prime}) \equiv \frac{1}{2} \{
   G[|{\bf r-r}^{\prime}|, \bar{n}({\bf r})] + 
   G[|{\bf r-r}^{\prime}|, \bar{n}({\bf r}^{\prime})]\}, 
\label{eq4.6}
\end{equation}
and  
\begin{equation}
G^{\rm SWDA}({\bf r}, {\bf r}^{\prime}) =
   G^{\rm SWDA}({\bf r}^{\prime}, {\bf r}).
\label{eq4.7}
\end{equation}
Now we have $v_{1}({\bf r}) = v_{2}({\bf r}) = \epsilon _{\rm xc} ({\bf r})$, and
the conditions \ref{eq4.1} and \ref{eq4.3} are fulfilled naturally.

The corresponding sum rule will be
\begin{equation}
\int n({\bf r}^{\prime}) 
G^{\rm SWDA}({\bf r}, {\bf r}^{\prime}) d{\bf r}^{\prime} =
\frac{1}{2} \int n({\bf r}^{\prime}) \{
G[|{\bf r-r}^{\prime}|, \bar{n}({\bf r})] + 
G[|{\bf r-r}^{\prime}|, \bar{n}({\bf r}^{\prime})]\} 
d{\bf r}^{\prime} = -1.
\label{eq4.8}
\end{equation}
Self-consistent iteration can be used to determine $\bar{n}({\bf r})$. Once 
$\bar{n}({\bf r})$ is known, $E^{\rm SWDA}_{\rm xc}$ is known also. The difficulty
arises when one wants to calculate $v_{3}({\bf r})$. If it can be determined 
efficiently, this symmetrized WDA will satisfy all the conditions listed in 
the beginning of last subsection, so this
new WDA is very promising. It would overcome the problems that lead to
shell partitioning because core and valence states would be distinguishable
to screen valence electrons. 

\section{Conclusions}
In conclusion, we calculated ground state properties of some common ferroelectric
perovskites with the WDA. Compared with results of the LDA and GGA, the WDA describes
properties of systems with the cubic symmetry or at experimental ferroelectric lattices 
very well, but it fails to predict good fully optimized structure. The symmetry 
problem of function $G$ could cause this failure. A new approach is proposed,
and efforts must be taken to circumvent the mathematical difficulty.

\section{Acknowledgments}
This work was supported by the Office of Naval Research under ONR Grants
N00014-02-1-0506 and N0001403WX20028. Work at NRL is supported by the Office
of Naval Research. Calculations were done on the Center for Piezoelectrics 
by Design (CPD) computer facility, and on the Cray SV1 supported by NSF 
and the Keck Foundation. We are grateful for discussions with H. Krakauer, M. Pederson,
I. Mazin, D. Vanderbilt, N. Choudhury, A. Asthagiri, M. Sepliarsky, M. Gupta and 
R. Gupta. We also thank E. Walter and M. Barnes for helping
us to smooth out computations on the CPD cluster.


\pagebreak

\begin{table}
\caption{Fully relaxed structures of tetragonal PbTiO$_3$ with the LDA and the GGA. 
Volumes are in \AA$^3$, the numbers in parentheses are the deviations
of strain and volume from experiment, and the experimental $c/a = 1.063$. 
\label{tab1}}
\begin{ruledtabular}
\begin{tabular}{l|cc|cc}

           &  volume (Expt.) & $c/a$ &  volume (Relaxed) & $c/a$       \\  
\hline
LDA        &  63.28  &  1.112 (+80\%) &  60.36 (-4.6\%) & 1.051 (-20\%)  \\
GGA        &  63.28  &  1.068 (+7\%)  &  70.58 (+11\%)  & 1.23  (+260\%) \\

\end{tabular}
\end{ruledtabular}
\end{table}


\begin{table}
\caption{Calculated LDA, GGA and WDA lattice constants in \AA \ for these ferroelectric 
materials in cubic state, compared with experimental data. Uniform electron gas $G$
was used in the WDA. Numbers in parentheses are deviations from experiment. 
\label{tab2}}
\begin{ruledtabular}
\begin{tabular}{l|ccc|c}

     material  & LDA          & GGA        & WDA         &  \ \ \ \ Expt. \ \ \ \ \\  
\hline					     
     KNbO$_3$  \ \ \ \  \ \ \ \  \ \ \ \     
               & 3.960 (-1.4\%) & 4.018 (+0.1\%) & 4.011 (-0.1\%) &  4.016  \\
     KTaO$_3$  & 3.931 (-1.3\%) & 4.032 (+1.2\%) & 3.972 (-0.3\%) &  3.983  \\
     SrTiO$_3$ & 3.858 (-1.2\%) & 3.935 (+0.8\%) & 3.917 (+0.3\%) &  3.905  \\
     BaTiO$_3$ & 3.951 (-1.4\%) & 4.023 (+0.6\%) & 4.009 (+0.2\%) &  4.000  \\
     PbTiO$_3$ & 3.894 (-1.9\%) & 3.971 (+0.1\%) & 3.941 (-0.7\%) &  3.969  \\

\end{tabular}
\end{ruledtabular}
\end{table}


\begin{table}
\caption{Optimized internal coordinates of tetragonal PbTiO$_3$ at the experimental volume
($V=63.28$ \AA$^3$) and strain ($c/a=1.063$). $u_z$ are given in terms of the lattice 
constant $c$. Uniform electron gas $G$ was used in the WDA. \label{tab3}}
\begin{ruledtabular}
\begin{tabular}{l|ccc|c}

                        & LDA   & GGA      & WDA  &  \ \ \ \ Expt. \ \ \ \ \\ 
\hline
     $u_z$(Pb)          & 0.000 & 0.000    & 0.000        & 0.000  \\
     $u_z$(Ti)          & 0.539 & 0.532    & 0.539        & 0.538  \\
     $u_z$(O$_1$,O$_2$) & 0.615 & 0.611    & 0.614        & 0.612  \\
     $u_z$(O$_3$)       & 0.111 & 0.105    & 0.110        & 0.117  \\

\end{tabular}
\end{ruledtabular}
\end{table}


\begin{table}
\caption{Optimized internal coordinates of rhombohedral BaTiO$_3$ for the experimental volume
($V=64.00$ \AA$^3$). $u_z$ are given in terms of the lattice constant. Uniform electron gas 
$G$ was used in the WDA. 
\label{tab4}}
\begin{ruledtabular}
\begin{tabular}{l|ccc|c}

                        & LDA   & GGA   & WDA  &  \ \ \ \ Expt. \ \ \ \ \\ 
\hline
     $u_z$(Ba)          & 0.000 & 0.000 & 0.000        & 0.000  \\
     $u_z$(Ti)          & 0.488 & 0.488 & 0.489        & 0.489  \\
     $u_z$(O$_1$,O$_2$) & 0.511 & 0.510 & 0.509        & 0.511  \\
     $u_z$(O$_3$)       & 0.020 & 0.018 & 0.017        & 0.018  \\

\end{tabular}
\end{ruledtabular}
\end{table}


\begin{table}
\caption{Optimized internal coordinates of rhombohedral KNbO$_3$ for the experimental volume
($V=64.77$ \AA$^3$). $u_z$ are given in terms of the lattice constant. 
Uniform electron gas $G$ was used in the WDA.
\label{tab5}}
\begin{ruledtabular}
\begin{tabular}{l|ccc|c}

                        & LDA   & GGA       & WDA  &  \ \ \ \ Expt. \ \ \ \ \\ 
\hline
     $u_z$(K)           & 0.507 & 0.509     & 0.508        & 0.5131  \\
     $u_z$(Nb)          & 0.000 & 0.000     & 0.000        & 0.0000  \\
     $u_z$(O$_1$,O$_2$) & 0.019 & 0.024     & 0.017        & 0.0313  \\
     $u_z$(O$_3$)       & 0.527 & 0.522     & 0.520        & 0.5313  \\

\end{tabular}
\end{ruledtabular}
\end{table}


\begin{table}
\caption{Volumes (\AA$^3$) and bulk modulus (GPa) of fully optimized tetragonal 
PbTiO$_3$, rhombohedral BaTiO$_3$ and KNbO$_3$. Uniform electron gas $G$ was used 
in WDA. Numbers in parentheses are deviations from experiment. Experimental data are 
from Refs. \cite{berlin}.
\label{tab6}}
\begin{ruledtabular}
\begin{tabular}{l|cccc|c}

           & & LDA          & GGA          & WDA  &  \ \ \ \ Expt. \ \ \ \ \\ 
\hline
 PbTiO$_3$ & V & 60.36 (-4.6\%) & 70.58 (+11\%)  & 68.46 (+8.2\%) & 63.28  \\
           & B & 83             & 40             & 42             & 204 \\
 BaTiO$_3$ & V & 61.59 (-3.8\%) & 67.47 (+5.4\%) & 66.82 (+4.4\%) & 64.00  \\
           & B & 148            & 94             & 99             & 196, 156, 139\\
 KNbO$_3$  & V & 61.96 (-4.3\%) & 66.63 (+2.9\%) & 66.04 (+2.0\%) & 64.77  \\
           & B & 163            & 98             & 114            & 138 \\
\end{tabular}
\end{ruledtabular}
\end{table}


\begin{table}
\caption{Calculated WDA (with four types of $G$ denoted in subsection E) lattice 
constants in \AA \ for these ferroelectric materials in cubic state. Numbers in 
parentheses are deviations from experiment. 
\label{tab7}}
\begin{ruledtabular}
\begin{tabular}{l|cccc}

     material  &  WDA (a)  &  WDA (b)  &  WDA (c) & WDA (d) \\  
\hline					     
     KNbO$_3$  \ \ \ \  \ \ \ \  \ \ \ \     
               & 4.011 (-0.1\%) & 4.018 (+0.0\%) & 4.006 (-0.2\%) & 4.001 (-0.4\%)    \\
     KTaO$_3$  & 3.972 (-0.3\%) & 3.979 (-0.1\%) & 3.969 (-0.3\%) & 3.964 (-0.5\%)    \\
     SrTiO$_3$ & 3.917 (+0.3\%) & 3.924 (+0.5\%) & 3.911 (+0.2\%) & 3.908 (+0.1\%)    \\
     BaTiO$_3$ & 4.009 (+0.2\%) & 4.015 (+0.4\%) & 4.011 (+0.3\%) & 4.004 (+0.1\%)    \\
     PbTiO$_3$ & 3.941 (-0.7\%) & 3.948 (-0.5\%) & 3.938 (-0.8\%) & 3.932 (-0.9\%)    \\

\end{tabular}
\end{ruledtabular}
\end{table}


\begin{table}
\caption{Volumes of fully optimized tetragonal PbTiO$_3$, rhombohedral 
BaTiO$_3$ and KNbO$_3$ in \AA$^3$ using the WDA with two types of
$G$. Numbers in parentheses are deviations from experiment. 
\label{tab8}}
\begin{ruledtabular}
\begin{tabular}{l|cc}

            & WDA (a)  & WDA (d) \\ 
\hline
 PbTiO$_3$  & 68.46 (+8.2\%) & 63.79 (+0.8\%)   \\
 BaTiO$_3$  & 66.82 (+4.4\%) & 65.55 (+2.4\%)  \\
 KNbO$_3$   & 66.04 (+2.0\%) & 65.60 (+1.3\%)  \\

\end{tabular}
\end{ruledtabular}
\end{table}


\begin{references}

\bibitem{cohen99}
See e.g., R.E. Cohen, J. Phys. Chem. Solids {\bf 61}, 139-146 (1999).

\bibitem{cohen90}
R.E. Cohen, and H. Krakauer, Phys. Rev. B {\bf 42}, 6416 (1990);
R.E. Cohen, Nature (London) {\bf 358}, 136 (1992);
R.E. Cohen, and H. Krakauer, Ferroelectrics {\bf 136}, 65 (1992).

\bibitem{singh92}
D.J. Singh, and L.L. Boyer, Ferroelectrics {\bf 136}, 95 (1992).

\bibitem{postn93}
A.V. Postnikov, T. Neumann, G. Borstel, and M. Methfessel,
Phys. Rev. B {\bf 48}, 5910 (1993);
A.V. Postnikov, T. Neumann, and G. Borstel, Ferroelectrics {\bf 164}, 101 (1995).

\bibitem{yu95}
R. Yu, and H. Krakauer, Phys. Rev. Lett. {\bf 74}, 4067 (1995);
C.-Z. Wang, R. Yu, and H. Krakauer, Ferroelectrics {\bf 194}, 97 (1997).

\bibitem{singh95}
D.J. Singh, Ferroelectrics {\bf 164}, 143 (1995);
Ferroelectrics {\bf 194}, 299 (1997).

\bibitem{corso94}
A. Dal Corso, S. Baroni, and R. Resta, Phys. Rev. B {\bf 49}, 5323 (1994).

\bibitem{resta97}
R. Resta, Ferroelectrics {\bf 194}, 1 (1997);
S. Dallolio, R. Dovesi, and R. Resta, Phys. Rev. B {\bf 56}, 10105 (1997).

\bibitem{hamann79}
D.R. Hamann, Phys. Rev. Lett. {\bf 42}, 662 (1979).

\bibitem{yin82}
M.T. Yin, and M.L. Cohen, Phys. Rev. B {\bf 26}, 5668 (1982).

\bibitem{lang83}
D.C. Langreth, and M.J. Mehl, Phys. Rev. B {\bf 28}, 1809 (1983);
A.D. Becke, Phys. Rev. A {\bf 38}, 3098 (1988).

\bibitem{perdew96}
J.P. Perdew, K. Burke, and M. Ernzerhof, Phys. Rev. Lett. {\bf 77}, 3865 (1996).

\bibitem{perdew92a}
J.P. Perdew, J.A. Chevary, S.H. Vosko, K.A. Jackson, M.R. Pederson, D.J. Singh, 
and C. Fiolhais, Phys. Rev. B {\bf 46}, 6671 (1992); 
Phys. Rev. B {\bf 48}, 4978(E) (1993). 

\bibitem{hammer93}
B. Hammer, K.W. Jacobsen, and J.K. Norskov, Phys. Rev. Lett. {\bf 70}, 3971 (1993); 
B. Hammer, and M. Scheffler, Phys. Rev. Lett. {\bf 74}, 3487 (1995).

\bibitem{filip94}
C. Filippi, D.J. Singh, and C.J. Umrigar, Phys. Rev. B {\bf 50}, 14947 (1994).

\bibitem{hedin71}
L. Hedin, and B.I. Lundqvist, I. Phys. C {\bf 4}, 2064 (1971).

\bibitem{singh94}
D.J. Singh, {\it Planewaves, Pseudopotentials and LAPW method}
(Kluwer Academic publishers, Boston, 1994).

\bibitem{abinit1}
X. Gonze, J.-M. Beuken, R. Caracas, F. Detraux, M. Fuchs, G.-M. Rignanese, 
L. Sindic, M. Verstraete, G. Zerah, F. Jollet, M. Torrent, A. Roy, M. Mikami, 
Ph. Ghosez, J.-Y. Raty, D.C. Allan, Comput. Mater. Sci. {\bf 25}, 478 (2002). 

\bibitem{abinit2}
The ABINIT code is a common project of the Université Catholique de Louvain, Corning 
Incorporated, and other contributors (URL http://www.abinit.org).

\bibitem{gunna76}  
O. Gunnarsson, M. Jonson, and B.I. Lundquist, Phys. Lett. {\bf 59A}, 177 (1976);
O. Gunnarsson, M. Jonson, and B.I. Lundquist, Solid State Commun. {\bf 24}, 765 (1977).

\bibitem{alonso78}  
J.A. Alonso, and L.A. Girifalco, Phys. Rev. B {\bf 17}, 3735 (1978).

\bibitem{gunna79}
O. Gunnarsson, M. Jonson, and B.I. Lundqvist, Phys. Rev. B {\bf 20}, 3136 (1979).

\bibitem{zunger81}
J.P. Perdew and A. Zunger, Phys. Rev. B {\bf 23}, 5048 (1981).

\bibitem{svane90}
A. Svane, and O. Gunnarsson, Phys. Rev. Lett. {\bf 65}, 1148 (1990);
Z. Szotek, W.M. Temmerman, and H. Winter, Phys. Rev. B {\bf 47}, 4029 (1993).

\bibitem{grit93}
O.V. Gritsenko, A. Rubio, L.C. Balb\'{a}s, and J.A. Alonso, 
Chem. Phys. Lett. {\bf 205}, 348 (1993).

\bibitem{singh93}
D.J. Singh, Phys. Rev. B {\bf 48}, 14099 (1993).

\bibitem{singh96}
D.J. Singh, Phys. Rev. B {\bf 53}, 176 (1996).

\bibitem{lasota}
C. Lasota, C. Wang, R. Yu, and H. Krakauer, Ferroelectrics {\bf 194} 109 (1997).

\bibitem{gunna1976}
O. Gunnarsson, and B.I. Lundqvist, Phys. Rev. B {\bf 13}, 4274 (1976).

\bibitem{perdew92b}
J.P. Perdew, and Yue Wang, Phys. Rev. B {\bf 46}, 12947 (1992).

\bibitem{gunna80}
O. Gunnarsson, and R.O. Jones, Phys. Scr. {\bf 21}, 394 (1980).

\bibitem{mazin98}
I.I. Mazin, and D.J. Singh, Phys. Rev. B {\bf 57}, 6879 (1998).

\bibitem{rush2002}
P.P. Rushton, D.J. Tozer, and S.J. Clark, Phys. Rev. B {\bf 65}, 193106 (2002).

\bibitem{wu2003}
Z. Wu, R.E. Cohen and D.J. Singh, in AIP conference proceedings volume {\bf 677} of 
{\it Fundamental Physics of Ferroelectrics} (2003), p. 276, 
Edited by P.K. Davies and D.J. Singh.

\bibitem{wu2004}
Z. Wu, R.E. Cohen, D.J. Singh, R. Gupta, and M. Gupta, 
Phys. Rev. B {\bf 69}, 085104 (2004).

\bibitem{hf37}
H. Hellmann, {\it Einfuhrung in die Quantenchemie} (Deuieke, Leipzig, 1973), p.
285; R.P. Feynman, Phys. Rev. {\bf 56}, 340 (1939).

\bibitem{tm91}
N. Troullier, and J.L. Martins, Phys. Rev. B {\bf 43}, 1993 (1991).

\bibitem{ggp02}
P. Gori-Giorgi and J.P. Perdew, Phys. Rev. B {\bf 66}, 165118 (2002).

\bibitem{hewat}
A.W. Hewat, J. Phys. C {\bf 6}, 2559 (1973).

\bibitem{berlin}
G.J. Fischer, W.C. Wang, and S. Karato, Phys. Chem. Minerals, {\bf 20}, 97 (1993).
D.A. Berlincourt, H. Jaffe, Phys. Rev {\bf 111}, 143 (1985).
L.R. Edwards, R.W. Lynch, J. Phys. Chem. sol., {\bf 31}, 573 (1970).
E. Wiesendanger, Ferroelectrics {\bf 6}, 263 (1974).

\end{references}
\end{document}